\documentclass{ws-ijmpb}
\usepackage{graphicx}
\newcommand{\mrm}[1]{\mathrm{#1}}
\newcommand{\mbf}[1]{\mathbf{#1}}

\begin{document}

\markboth{GREGOR WEIHS}{PHOTONIC CRYSTAL WAVEGUIDES FOR PARAMETRIC
DOWN-CONVERSION}

\catchline{}{}{}{}{}

\title{PHOTONIC CRYSTAL WAVEGUIDES FOR PARAMETRIC DOWN-CONVERSION}

\author{GREGOR WEIHS}
\address{Institute for Quantum Computing and Department of Physics,
University of Waterloo, 200 University Ave W, Waterloo, Ontario, N2L
3G1, Canada \\ weihs@iqc.ca}

\maketitle

\begin{history}
  \received{07 10 2005}
\end{history}

\begin{abstract}
Photonic crystals create dramatic new possibilities for nonlinear
optics. Line defects are shown to support modes suitable for the
production of pairs of photons by the material's second order
nonlinearity even if the phase-matching conditions cannot be
satisfied in the bulk. These structures offer the flexibility to
achieve specific dispersion characteristics and potentially very
high brightness. In this work, two phase matching schemes are
identified and analyzed regarding their dispersive properties.
\end{abstract}


\section{INTRODUCTION}
For many applications there is a need for new sources of entangled
photon pairs that are brighter or show certain dispersion
characteristics. In quantum communication for example, it is
entanglement based quantum key distribution, quantum teleportation
and, most importantly, quantum repeaters that all depend on a good
supply of entangled photon pairs.

But photon pairs have more traditional applications as well. Phase
measurements and lithography may one day employ entanglement to beat
their classical limits. In these scenarios one would not only look
for entangled pairs, but possibly even for entangled states of
multiple photons. At present, though, the efforts to even create
enough pairs to begin with, far outweighs the possible gains in
precision. It is yet more difficult and costly in terms of resources
to interferometrically engineer higher entangled states from pairs.

This would not necessarily be the case if only there were sources
available that could be plugged into a wall outlet and had fiber
outputs providing the entangled pairs, or even higher dimensional
entangled photon states. To date the majority of experiments obtains
their pairs from parametric down-conversion in bulk
materials\cite{Kwiat95b}.

One standard way of achieving nonlinear interaction in materials
that are not a priori suitable for this purpose is to use periodic
poling. In periodic poling the sign of the nonlinear tensor
component is reversed periodically. However, the linear dispersion
characteristics all stay the same. Still, periodically poled
crystals and waveguides have been used as high-yield sources of
entangled photon pairs\cite{Sanaka01a,Tanzilli01a}.

Photonic crystals allow us to go beyond the properties of natural
materials in many ways\cite{Joannopoulos95a}. They have been shown
to exhibit super strong or very small dispersion, enhance nonlinear
interactions\cite{Cowan05a}, and to confine waves in guides and
resonators. Therefore it only seems natural to consider photonic
crystals for the production of entangled photon
pairs\cite{Vamivakas04a}. De Dood \textit{et al.}\cite{deDood04a}
recently suggested to exploit form birefringence in multilayer
stacks (one-dimensional photonic crystals) for achieving
phase-matching in GaAs.

\section{PARAMETRIC DOWN-CONVERSION IN PHOTONIC CRYSTAL DEFECT
WAVEGUIDES}

Spontaneous parametric down-conversion is difference-frequency
generation where a pump beam at angular frequency $\omega_p$ and
wavevector $\mbf{k}_p$ irradiates a material with some nonlinear
tensor $\chi^{(2)}$. Then there is a probability to create pairs of
photons at $(\omega_1, \mbf{k}_1)$ and $(\omega_2, \mbf{k}_2)$ so
that $\omega_1+\omega_2=\omega_p$ (\emph{energy conservation}) and
$\mbf{k}_1+\mbf{k}_2=\mbf{k}_p$ (\emph{phase-matching}). In some
cases phase-matching can even be arranged such that the photon pairs
are polarization entangled\cite{Kwiat95b}.

Obviously, $\chi^{(2)}$ should be large and one should choose the
largest component of the tensor by selecting the polarizations and
directions of the involved light fields to maximize the effective
nonlinearity. In bulk materials, however, it is not always possible
to work in the maximizing configuration because the natural
dispersion of the material requires one to choose certain directions
in order to satisfy the phase-matching condition.

Periodicity can come to the rescue. In a periodic structure we know
from Bloch's theorem that the solutions of Maxwell's equation will
be periodic functions multiplied by a plane wave. The periodic
functions are labeled by a wavevector $\mbf{k}$ and we only need to
consider wavevectors within the first Brillouin zone. This means
that the phase-matching condition now reads
$\mbf{k}_1+\mbf{k}_2=\mbf{k}_p+\mbf{G}$ where $\mbf{G}$ is any
wavevector of the reciprocal lattice. In other words, any wavevector
we consider can be mapped to one in the first Brillouin zone and
phase-matching has to be satisfied within this zone.

Periodicity has been used in the form of periodic poling. In certain
crystals such as LiNbO$_3$ and KTiOPO$_4$ it is possible to reverse
the sign of the nonlinear optic coefficient. This will make it
possible to achieve phase-matching in otherwise forbidden
interaction schemes and therefore to utilize the maximum
nonlinearity and a long interaction region. Other materials that are
promising because of their large nonlinearity, such as GaAs, cannot
easily be poled, even though some progress has been
made\cite{Skauli02a}.

Because the number of pairs created in parametric down-conversion is
in the low excitation regime linear with the applied pump power
there is usually no need to go to pulsed sources. However, for
applications that need more than one photon pair at a time, such as
entanglement swapping, one needs to down-convert from ultra-fast
pulses\cite{Zukowski93a}. In this case, in addition to the
phase-matching condition one should ideally achieve group matching
as well. Otherwise there will be some differential group delay
between the pump and converted waves, or even between the two
converted waves if they have different polarization. In the source
of entangled photon pairs first described by Kwiat \textit{et
al.}\cite{Kwiat95b} differential group delay severely limits the
achievable entanglement and/or brightness, when it is used with a
fast pulsed pump and not limited to a very short interaction
region\cite{Keller97a}.

\begin{figure}
  \centerline{\includegraphics[width=0.3\textwidth]{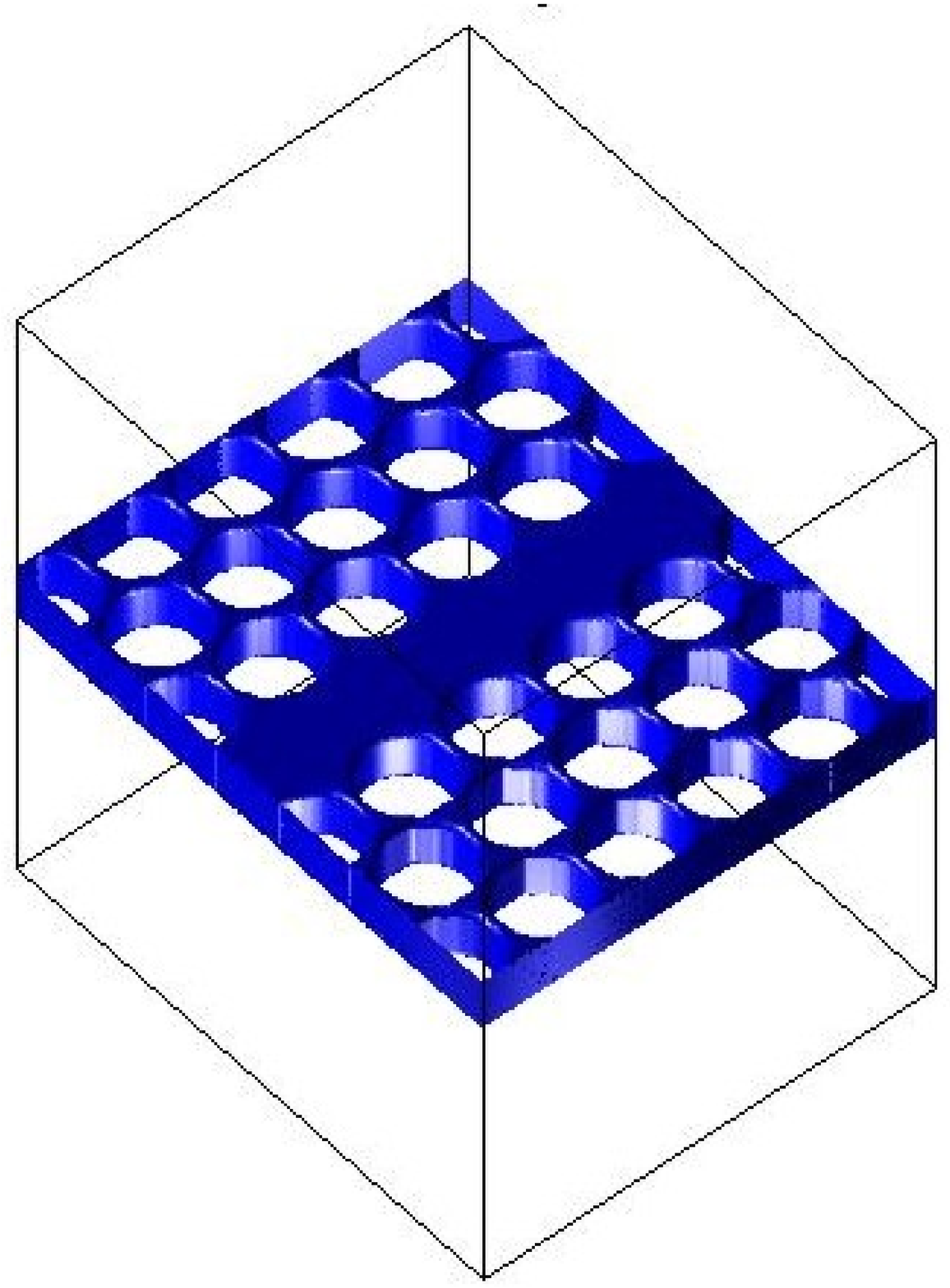}
  \includegraphics[width=0.6\textwidth]{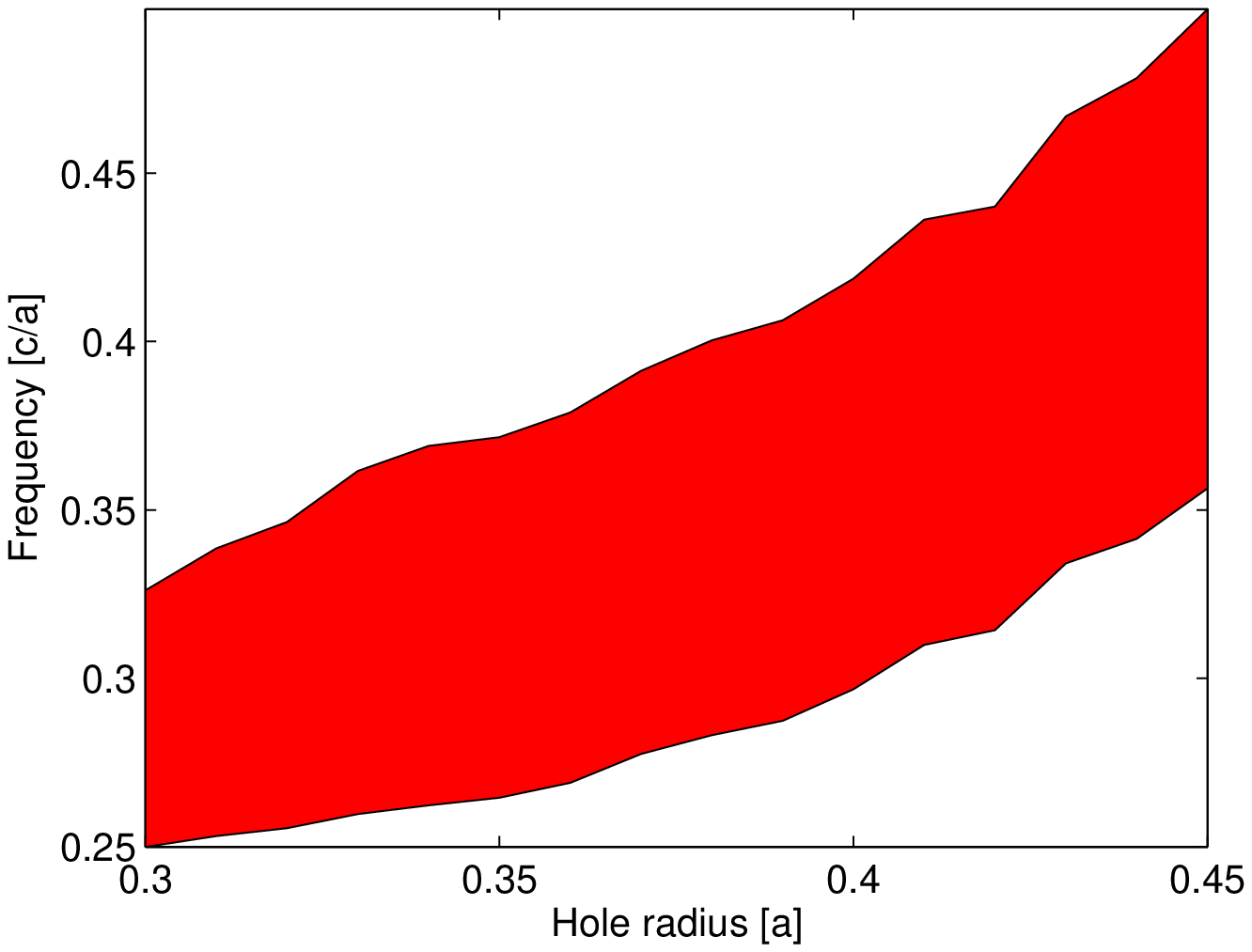}}
  \caption{\textbf{(Left)} A membrane of GaAs with a hexagonal array of holes except
  for one missing row. The missing row forms an increased-index
  waveguide. \textbf{(Right)} The shaded area represents the lowest lying gap in the even
  optical bands within the region outside the light cone for a slab of material with dielectric
  constant $\epsilon=13$ patterned with an hexagonal array of holes with
  lattice constant $a$. One sees that the bottom of the gap increases
  considerably with the hole size. It is important to remember that confinement
  perpendicular to the plane of the slab is only due to total internal reflection.
  Therefore the bandgaps are not true gaps, but rather gaps for light propagating in the slab.}
  \label{fig1}
\end{figure}

While periodic poling solves the phase-matching problem, it does not
change the group velocities, because the effective refractive index
is not changed. A natural consequence is therefore to not only
modulate the nonlinearity but the dielectric as a whole, i.e. to
look at photonic crystal structures. In contrast to earlier
suggestions\cite{Vamivakas04a,deDood04a}, here we will concentrate
on waveguide structures that are formed by a line defect in a
slab-type photonic crystal as shown in Fig.~\ref{fig1}. The purpose
of having a waveguide is a long interaction region and good mode
overlap.

Waveguides in photonic crystal slabs (PCS) have first been studied
theoretically by Johnson \textit{et al.}\cite{Johnson00a}. There are
many possibilities to create waveguides in photonic crystals. The
most popular configuration is a line defect in a lattice of
air-holes produced by reactive ion etching with subsequent undercut
so as to create a free-standing membrane of materials such as GaAs,
InP or Si. Spectacular structures up to 1~cm long have been
demonstrated with losses as low as 0.76~dB/cm \citelow{Sugimoto04a}.

AlGaAs is a favorable material from many points of view. The fact
that its lattice constant is almost independent of the Aluminum
fraction makes it possible to grow arbitrary heterostructures. Thus,
the fundamental electronic band gap can be chosen freely. It has
very high refractive index varying from about 3.6 (for 15\%
Aluminium) at 775~nm to about 3.3 at 1550~nm. The reported nonlinear
susceptibilities scatter but in recent years a consensus seems to
have been reached\cite{Skauli02a} at values around 100~pm/V, which
is about 5 times higher than for LiNbO$_3$ and almost 50 times that
of $\beta$-BaB$_2$O$_4$. The maximum effective nonlinearity is
achieved for three identical polarizations parallel to the [111]
direction. Further, if any of the interacting waves is polarized
along [011] the effective nonlinearity is independent of the other
waves' polarization. Growth of AlGaAs is usually done on [100] cut
wafers. Light propagating in a membrane parallel to the surface
could therefore propagate in any direction orthogonal to [100].

Fully three-dimensional band structure calculations using the MIT
Photonic Bands package\cite{Johnson01a} helped identify possible
phase-matching schemes. In the literature there is very little data
on band gaps for PCS as a function of either hole radius, or slab
thickness or dielectric constant. Since for the problem at hand we
need to be able to choose the location of the band gap with some
care, I first calculated a series of bandstructures resulting in a
map of the gap locations as a function of the hole radius. The
shaded area in Fig.~\ref{fig1}(right) shows the bandgap for even
modes in an hexagonal array of holes patterned into a membrane as a
function of the hole radius.

A photonic crystal defect waveguide in which the defect has
increased refractive index with respect to the periodic structure
around it can support two types of guided modes. There are
index-guided modes lying below the lowest PCS bands and modes that
lie within the band gap of the PCS. Obviously, the confinement will
be better for the latter, but if there are no sharp bends, index
guided modes should not suffer from much higher losses. Further, in
a PCS that is symmetric about its central plane the bands split into
even and odd ones, closely corresponding to their TE and TM
polarized counterparts in two-dimensional photonic crystals.

The allowed regions for defect waveguide modes are determined by
projecting the bands of the perfect PCS onto the guiding
k-direction. This procedure masks all areas in the $(\omega,k_x)$
strip that are covered by states that are extended in the plane and
can thus not support modes that would be localized to a defect. This
is in addition to the restriction that all guided modes lie outside
the lightcone of the surrounding medium. In the cases investigated
here, the surrounding medium is air around the free-standing
Al$_{0.15}$Ga$_{0.85}$As membrane. The material choice is mainly
motivated by the desire to create photon pairs at telecommunications
wavelengths, i.e. 1550~nm. The pump would then have to be at 775~nm
for a symmetric source which requires a 10\% Aluminium fraction or
higher. At 15\% one should be safely outside all exciton and
impurity resonances and expect very little absorption for the pump.

From studying gap maps for hexagonal and square lattices of holes,
we can conclude that either may be suitable for waveguiding, with
the hexagonal ones yielding larger gaps and thus possibly more
flexibility. The most basic defect waveguide is a single missing row
of holes (see Fig.~\ref{fig1}), in which waves then propagate in the
$\Gamma$-K direction of the perfect PCS. It is well
known\cite{Johnson00a} that such a waveguide is not single-mode.
Single-mode behavior can be brought about by decreasing the width of
the defect\cite{Benisty96a}. It is clear that for a proper source,
suitable for the generation of entangled pairs single mode behavior
will be essential.

\begin{figure}
  \centerline{\includegraphics[width=0.6\textwidth]{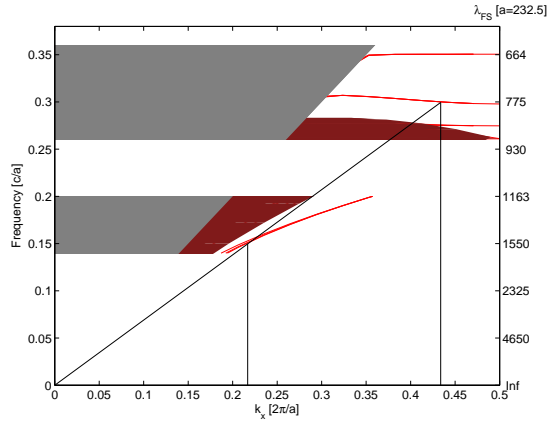}}
  \caption{Dispersion diagram (frequency vs. $k_x$) for the z-even
  modes of a structure like in Fig.~\protect\ref{fig1} with the
  projected bands of the PCS and the lightcone masked out. The hole
  radius is 0.38$a$, the dielectric constant is 13.0 for the upper
  section and 10.8 for the lower one. The close to
  horizontal modes in the upper section are gap-guided defect modes,
  whereas in the lower section we only find index-guided modes. The
  straight line indicates a potential phase-matching scheme at the
  points where it intersects with the waveguide modes. For easy
  reference a (vacuum) wavelength scale is plotted assuming a
  lattice constant of $a=232.5$~nm.} \label{eveneven}
\end{figure}

\begin{figure}
  \centerline{\includegraphics[width=0.6\textwidth]{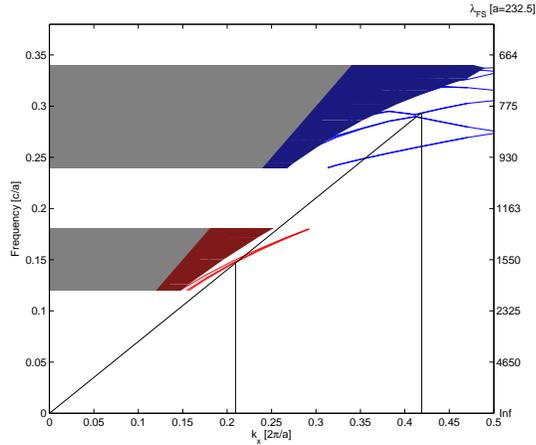}}
  \caption{Similar to Fig.~\ref{eveneven}, except that the upper
  section now shows the z-odd bands instead of the z-even ones. A line
  again connects points that match $k_p=k_1+k_2$ and
  $\omega_p=\omega_1+\omega_2$.}
  \label{oddeven}
\end{figure}

Figure~\ref{eveneven} shows a scenario in which photons from a
z-even gap-guided mode can down-convert into two photons of a z-even
index-guided band.\protect\footnote{Unfortunately the frequency
domain method used in the MIT Photonic Bands package cannot treat
dispersive materials. Therefore the calculation was split into two,
one for the low refractive index at long wavelengths, and one for
the high refractive index and shorter wavelength.}
Figure~\ref{oddeven} on the other hand shows a configuration in
which pump photons in a z-odd index-guided band down-convert into
photon pairs of a z-even index-guided band. The lattice is
responsible for producing such a vastly different dispersion for the
two different symmetries\cite{Joannopoulos95a}.

In any case, the relative strength of the nonlinear interaction will
depend on the overlap of the three fields involved. The amplitude of
the one photon pair term in the output state will be proportional to
(in first order perturbation theory)
\[  \int_V d^3\mathbf{r} \chi^{(2)}_{ijk} E_i E_j E_k, \]
where $E_{i,j,k}$ are three components of the electrical field and
we'll have to take the sum over all indices. One of the field
components would be the pump (high frequency) and the two other ones
the down-conversion fields (low frequency). For the case where the
low-frequency modes are degenerate the integrand reduces to $E_i
E_j^2$. In the case of a periodic structure, we can take the
integral over a unit cell. Obviously, the field distribution within
the unit cell has to be similar between the three fields to achieve
substantial overlap. Also, the fields will only contribute within
the material, but not in the air space within the holes or above and
below the slab. Preliminary calculations show that at least the
phase-matching scheme shown in Fig.~\ref{eveneven} achieves
reasonable overlap for band number 9  (counted from 0 frequency) at
$k=0.44$ with band number 1 at $k=0.22$. It is part of our ongoing
work to calculate the effective nonlinearity in absolute units.

\section{DISCUSSION}

The goal for parametric down-conversion in photonic crystals is to
exploit the high nonlinearity of certain semiconductors and possibly
achieve group velocity matching. Group velocity matching is similar
to achieving phase-matching over an extended bandwidth, if the band
curvature is not too big.

It is obvious that the above schemes do not yet achieve
group-matching, where the second one is closer to the goal but still
too far away to be practically relevant for this purpose. Given that
the gap-guided modes exhibit very small group velocities, it would
be exciting if there was a way to phase-match between bands of
different gaps. However, since the gaps in a PCS are typically
shifted up in frequency by the vertical confinement it seems very
difficult to find gaps at one frequency and twice the frequency
simultaneously, where both are also below the light cone in the
waveguiding direction.

For photon pairs created by ultrafast pulses one has to compare the
differential group delay (DGD) with the pulse duration. If the group
velocities are $u_\mrm{pump}$ and $u_\mrm{dc}$ then the DGD per
length is just $|1/u_\mrm{pump}-1/u_\mrm{dc}|$, which means that for
small group velocities it will be more difficult to achieve a small
DGD.

Yet, for some applications, such as interfacing to
electromagnetically induced transparency (EIT) and stored light it
is desirable to have very narrow-band down-conversion sources. For
this purpose various groups have considered counter-propagating
solutions, which are typically phase-matched only in a point, or
equivalently, have an extremely high DGD, because one of the
velocities is negative. For this purpose the first even-even-even
phase-matching scheme (Fig.~\ref{eveneven}) appears to be a very
good solution.

\section{CONCLUSIONS}

So far, we have only considered degenerate cases, i.e. both
down-conversion photons have the same frequency. Asymmetric sources
have applications in the preparation of single-photon states and in
schemes where one photon needs to be detected with high efficiency,
whereas the other needs to propagate with very low loss through an
optical fiber. Clearly, there are many solutions for non-degenerate
phase-matching to be found in the above diagrams. Whether a certain
solution is interesting depends on the details of the intended
application.

I have shown that phase matching is in principle possible in
photonic crystal waveguides. The coupling of both the pump light and
the down-converted one in and out of the waveguide respectively is a
challenge. Cleaved edges can be used\cite{Kramper04a}, as well as
fiber tapers\cite{Barclay03a}, where the latter have achieved the
highest reported coupling efficiency to date.

The biggest interest in photon pairs is associated with
entanglement. With a single waveguide mode and polarization there
can be no entanglement a priori. However, the photons of a pair are
still created simultaneously and therefore it should always be
possible to construct a source of time-bin entangled photon pairs as
is customary for photon pairs from waveguides\cite{Tanzilli01a}.
With a solution that creates pairs in two different polarizations
one can, using two identical waveguides, even construct a source of
polarization entangled photon pairs.

In conclusion, we have seen that it is possible to exploit the high
optical nonlinearity of AlGaAs and to achieve phase-matching in a
slab-type photonic crystal defect waveguide. The next steps are to
perform fully dispersive calculations and to tune the waveguide
modes by changing the waveguide's width and edge shape so as to give
the involved modes the desired dispersion. While some details are
still missing in this picture an ultra bright source of photon pairs
at telecommunication wavelengths based on photonic crystals seems
within reach.

\section*{Acknowledgements}

This work was supported by the Canadian Foundation for Innovation
(CFI), the Canadian Institute for Advanced Research (CIAR), the
National Science and Engineering Research Council (NSERC), the
Ontario Research and Development Challenge Fund (ORDCF), and the
Ontario Photonics Consortium (OPC). The author wishes to thank R.
Horn for challenging discussions, J. Racle for help with part of the
calculations and S. G. Johnson for help with his MIT Photonic Bands
package.

\bibliographystyle{osa}
\bibliography{C:/Gregor/Litera/qobib}

\end{document}